\documentclass[11pt]{amsart} 
\usepackage{amscd,amssymb,amsxtra}
\usepackage{mathrsfs}  
\DeclareSymbolFontAlphabet{\mathrsfs}{rsfs}
\usepackage[mathscr]{eucal} 
\usepackage{comment}
\usepackage{color}
\usepackage{enumitem} 
\usepackage[colorlinks,backref=page,citecolor=refkey]{hyperref}
\usepackage{aliascnt}

\usepackage{marginnote}

\makeatletter
\let\@secnumfont\bfseries
\def\section{\@startsection{section}{1}%
  \z@{4\linespacing\@plus\linespacing}{\linespacing}%
  {\bfseries\centering}}
\def\introsection{\@startsection{section}{1}%
  \z@{3\linespacing\@plus\linespacing}{\linespacing}%
  {\bfseries\centering}}
\def\subsection{\@startsection{subsection}{2}%
   \z@{1.25\linespacing\@plus.7\linespacing}{.5\linespacing}%
   {\normalfont\bfseries}}
\def\subsectionsinline{\def\subsection{\@startsection{subsection}{2}%
  \z@{1\linespacing\@plus.7\linespacing}{-.5em}%
  {\normalfont\bfseries}}}

\makeatother

\numberwithin{equation}{section}

\newcommand{\mynewtheorem}[2]{
  \newaliascnt{#1}{equation}
  \newtheorem{#1}[#1]{#2}
  \aliascntresetthe{#1}
  \expandafter\def\csname #1autorefname\endcsname{#2}
}

\theoremstyle{definition}
\mynewtheorem{definition}{Definition}
\mynewtheorem{example}{Example}
\mynewtheorem{problem}{\color{blue}Problem}
\mynewtheorem{probsec}{\color{blue}Problem}
\mynewtheorem{exercise}{Exercise}
\mynewtheorem{question}{\color{blue}Question}
\mynewtheorem{project}{\color{blue}Project}
\mynewtheorem{construction}{Construction}
\mynewtheorem{task}{\color{blue}Task}
\mynewtheorem{otask}{\color{green}Obsolete Task}
\mynewtheorem{notation}{Notation}

\newtheorem*{definition*}{Definition}
\newtheorem*{example*}{Example}
\newtheorem*{problem*}{\color{blue}Problem}
\newtheorem*{probsec*}{\color{blue}Problem}
\newtheorem*{exercise*}{Exercise}
\newtheorem*{question*}{\color{blue}Question}
\newtheorem*{project*}{\color{blue}Project}
\newtheorem*{construction*}{Construction}
\newtheorem*{notation*}{Notation}

\theoremstyle{remark}
\mynewtheorem{note}{Note}
\mynewtheorem{remark}{Remark}
\mynewtheorem{data}{Data}

\newtheorem*{note*}{Note}
\newtheorem*{remark*}{Remark}
\newtheorem*{data*}{Data}

\theoremstyle{plain}
\mynewtheorem{theorem}{Theorem}
\mynewtheorem{corollary}{Corollary}
\mynewtheorem{lemma}{Lemma}
\mynewtheorem{proposition}{Proposition}
\mynewtheorem{conjecture}{Conjecture}
\mynewtheorem{claim}{Claim}
\mynewtheorem{proposal}{Proposal}
\mynewtheorem{conclusion}{Conclusion}
\mynewtheorem{hypothesis}{Hypothesis}
\mynewtheorem{assumption}{Assumption}

\newtheorem*{theorem*}{Theorem}
\newtheorem*{corollary*}{Corollary}
\newtheorem*{lemma*}{Lemma}
\newtheorem*{proposition*}{Proposition}
\newtheorem*{conjecture*}{Conjecture}
\newtheorem*{claim*}{Claim}
\newtheorem*{proposal*}{Proposal}
\newtheorem*{conclusion*}{Conclusion}
\newtheorem*{hypothesis*}{Hypothesis}
\newtheorem*{assumption*}{Assumption}

\newenvironment{proof*}[1][\proofname]{
  \begin{proof}[#1]}{  
\end{proof}}

\definecolor{refkey}{rgb}{0,.6,.4}

\renewcommand{\:}{\colon}

\newcommand{\CC}{{\mathbb C}}

\DeclareMathOperator{\End}{End}

\DeclareMathOperator{\Hom}{Hom}
\DeclareMathOperator{\id}{id}

\DeclareMathOperator{\Map}{Map}

\DeclareMathOperator{\pt}{pt}
\newcommand{\QQ}{{\mathbb Q}}

\newcommand{\RR}{{\mathbb R}}

\newcommand{\ZZ}{{\mathbb Z}}

\newcommand{\chiup}{\raise.5ex\hbox{$\chi$}}
\newcommand{\cir}{S^1}

\newcommand{\inv}{^{-1}}
\DeclareRobustCommand{\mstrut}{^{\vphantom{1*\prime y\vee M}}}

\newcommand{\res}[1]{\negmedspace\bigm|\mstrut_{#1}}

\newcommand{\temsquare}{\raise3.5pt\hbox{\boxed{ }}}

\newcommand{\zmod}[1]{\ZZ/#1\ZZ}

\newcommand{\zt}{\zmod2}

\newcommand{\longhookrightarrow}{\lhook\joinrel\longrightarrow}
\newcommand{\hneg}{\mkern-.5\thinmuskip}
 
\DeclareFontFamily{U}{mathx}{}
\DeclareFontShape{U}{mathx}{m}{n}{<-> mathx10}{}
\DeclareSymbolFont{mathx}{U}{mathx}{m}{n}
\DeclareMathAccent{\widehat}{0}{mathx}{"70}
\DeclareMathAccent{\widecheck}{0}{mathx}{"71}
 
\DeclareMathSymbol{\bigtimes}{1}{mathx}{"91}

\newcommand{\upplus}{^{>0}}
\newcommand{\upnn}{^{\ge0}} 

\newcommand{\Znn}{\ZZ\upnn}
\newcommand{\Rp}{\RR\upplus}

\usepackage[all,2cell]{xy}\renewcommand{\cir}{\ensuremath{S^1}}
\usepackage{graphicx}\usepackage{epsf}
\definecolor{refkey}{rgb}{0,.8,.2}\definecolor{labelkey}{rgb}{1,0,0} 
\definecolor{cy}{rgb}{0,.8,.8}\definecolor{or}{rgb}{1,.55,0}

\DeclareMathOperator{\Bord}{Bord}
\DeclareMathOperator{\Cat}{Cat}
\DeclareMathOperator{\Edge}{Edge}
\DeclareMathOperator{\Fun}{Fun}
\DeclareMathOperator{\Vect}{Vect}
\DeclareMathOperator{\spec}{spec}
\newcommand{\FG}{F\mstrut _{\hneg G}}
\newcommand{\TC}{T\mstrut _{\hneg C}}
\newcommand{\bord}[1]{\Bord_{\langle#1\rangle}}
\newcommand{\dual}{^{\vee}}
\newcommand{\gpd}{/\!/} 

\newcommand{\sD}{\mathscr{D}}
\newcommand{\sF}{\mathscr{F}}
\newcommand{\sH}{\mathscr{H}}

\newcommand{\sO}{\mathcal{O}}
\newcommand{\tF}{\widetilde{F}_{\hneg G}}
\newcommand{\tL}{\widetilde{\Lambda  }}

\begin{document}

\abovedisplayskip18pt plus4.5pt minus9pt
\belowdisplayskip \abovedisplayskip
\abovedisplayshortskip0pt plus4.5pt
\belowdisplayshortskip10.5pt plus4.5pt minus6pt
\baselineskip=15 truept
\marginparwidth=55pt

\makeatletter
\renewcommand{\tocsection}[3]{%
  \indentlabel{\@ifempty{#2}{\hskip1.5em}{\ignorespaces#1 #2.\;\;}}#3}
\renewcommand{\tocsubsection}[3]{%
  \indentlabel{\@ifempty{#2}{\hskip 2.5em}{\hskip 2.5em\ignorespaces#1%
    #2.\;\;}}#3} 
\renewcommand{\tocsubsubsection}[3]{%
  \indentlabel{\@ifempty{#2}{\hskip 5.5em}{\hskip 5.5em\ignorespaces#1%
    #2.\;\;}}#3} 
\def\@makefnmark{%
  \leavevmode
  \raise.9ex\hbox{\fontsize\sf@size\z@\normalfont\tiny\@thefnmark}} 
\def\multfoot{\textsuperscript{\tiny\color{red},}}
\def\footref#1{$\textsuperscript{\tiny\ref{#1}}$}
\makeatother

\setcounter{tocdepth}{2}



 \title[Discrete QM systems and TFT]{Discrete quantum systems from topological field theory} 

 \author[D. S. Freed]{Daniel S.~Freed}
 \address{Harvard University \\ Department of Mathematics \\ Science Center
Room 325 \\ 1 Oxford Street \\ Cambridge, MA 02138}
 \email{dafr@math.harvard.edu}
 \thanks{DSF is supported by the Simons Foundation Award 888988 as part of the
Simons Collaboration on Global Categorical Symmetries.}

 \author[M. J. Hopkins]{Michael J.~Hopkins}
 \address{Harvard University \\ Department of Mathematics \\ Science Center
Room 325 \\ 1 Oxford Street \\ Cambridge, MA 02138}
 \email{mjh@math.harvard.edu}
 \thanks{MJH is supported by the Simons Foundation Award 888992 as part of the
Simons Collaboration on Global Categorical Symmetries.}

  \author[C. Teleman]{Constantin Teleman} 
  \address{Department of Mathematics \\ University of California \\ 970 Evans
 Hall \#3840 \\ Berkeley, CA 94720-3840}  
  \email{teleman@math.berkeley.edu}
  \thanks{CT is supported by the Simons Foundation Award 824143 as part of the
Simons Collaboration on Global Categorical Symmetries.}

  \thanks{This work was performed in part at Aspen Center for Physics (ACP),
which is supported by National Science Foundation grant PHY-2210452.  This
research was supported in part by grant NSF PHY-2309135 to the Kavli Institute
for Theoretical Physics (KITP)}

 \date{June 5, 2025}
 \begin{abstract} 
 We introduce a technique to construct gapped lattice models using defects in
topological field theory.  We illustrate with 2+1 dimensional models, for
example Chern--Simons theories.  These models are local, though the state space
is not necessarily a tensor product of vector spaces over the complex numbers.
The Hamiltonian is a sum of commuting projections.  We also give a topological
field theory construction of Levin--Wen models.
 \end{abstract}
\maketitle

In this note we share some constructions of discrete quantum mechanical systems
based on the mathematical foundation of topological field theory.  In
particular, we construct lattice systems that model 3-dimensional Chern-Simons
theories: they are gapped systems whose low energy limit is Chern-Simons
theory.  The methods we use apply in more generality and in any dimension, but
we make no general claims in this note.  Instead, we limit ourselves to
exposing a few examples.  In all of these models space is discrete, but time
can be discrete (\emph{stat mech models}) or continuous (\emph{lattice
models}).  In the latter case the relativistic mixing of space and time is
broken, whereas in the former case the mixing can remain.

A first theme of this note is that one should be flexible about the ``discrete
structure'' of space, which can take many forms.  A second theme is that one
should push the framework of discrete models beyond usual constructions.  A
third theme is that we can do so on firm foundations within the mathematically
well-developed setting of topological field theory.
 
Standard quantum field theories that appear in physics and in mathematical
applications begin as relativistic theories on Minkowski spacetime.  They
satisfy strong locality properties, of which we emphasize two.  First, in
Wick-rotated form they can be formulated on \emph{arbitrary} backgrounds:
smooth manifolds equipped with background fields, such as a Riemannian metric.
Second, the state space of an $n$-dimensional theory depends locally on an
$(n-1)$-manifold~$Y$ in the following sense.  If $Y=Y_1\cup \mstrut _{Z}Y_2$ is
expressed as a union of manifolds with boundary~$Z$, then there is a von
Neumann algebra attached to~$Z$; modules over it attached to~$Y_1,Y_2$; and the
state space of~$Y$ is the tensor product of those modules \emph{over the von
Neumann algebra}.\footnote{In precise versions of these statements, spacetime
and space are not cut sharply; rather, the manifolds~$Y_1,Y_2,Z$ are embedded
in germs of $n$-manifolds~\cite{Se,KS,We}} In topological field theory one is
accustomed to a similar locality of state spaces, though the algebra or
category attached to~$Z$ satisfies strong finiteness conditions.  The discrete
models we construct here exhibit both forms of locality: they can be placed on
general backgrounds and the state spaces depend locally on space with its
discrete structure.  In these models the latter locality can be written as a
tensor product over a finite dimensional algebra, which is generally not~$\CC$.

We begin with a brief description of the toric code in~\S\ref{sec:1}, told in
terms we need later on.  Even here we observe that the discrete structure of
space can be more general than usual: a CW~structure.  Novel ideas appear
in~\S\ref{sec:2}, where we explain how to build a local lattice model from a
3-dimensional topological field theory, such as Chern-Simons theory, and a
choice of a finite set of codimension two defects.  The models we build have
commuting projector Hamiltonians, which implies that they are gapped.  We
explain how to recover the toric code in this framework.  The Levin--Wen model
can also be realized using topological field theory with defects, as we explain
in~\S\ref{sec:4} following~\cite{KK}.  In~\S\ref{sec:3} we review some
highlights of our work~\cite{FT1} expressing the $1+1$ dimensional Ising
model---a stat mech model---in terms of topological field theory.
 
We thank the Aspen Center for Physics and the Kavli Institute for Theoretical
Physics for their hospitality and for providing a stimulating environment.
We also thank Mike Freedman, Alexei Kitaev, and Greg Moore for stimulating
discussions on a wide range of topics related to this work.

   \section{Prelude: the toric code}\label{sec:1}

The toric code was introduced by Alexei Kitaev~\cite{Ki}.  The model can be
defined for any finite group~$G$; here, for simplicity, we take~$G$ to be the
cyclic group of order two.

  \begin{figure}[ht]
  \centering
  \includegraphics[scale=1.6]{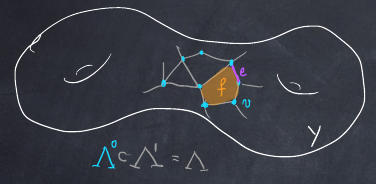}
  \vskip -.5pc
  \caption{A closed surface~$Y$ with an embedded graph~$\Lambda $}\label{fig:1}
  \end{figure}

Let $Y$~be a closed smooth 2-dimensional manifold, and suppose $\Lambda \subset
Y$ is a finite graph (or ``lattice'') that is smoothly embedded in~$Y$, as
depicted in \autoref{fig:1}.  We comment below (\autoref{thm:g21}) on a
generalization of this discrete structure on~$Y$.  Let $\Lambda ^0\subset
\Lambda $ denote the set of vertices of the graph~$\Lambda $.  Consider the
finite set
  \begin{equation}\label{eq:g46}
     \sD(\Lambda ,\Lambda ^0) = \left\{ \bigl(\tL\longrightarrow  \Lambda
     ,\,s\bigr): \tL\to \Lambda 
     \textnormal{ is a double cover},\, \textnormal{$s$ is a section of
     }\tL\res{\Lambda ^0}\longrightarrow  \Lambda ^0 \right\} \bigm/\;\cong 
  \end{equation}
of isomorphism classes of double covers of~$\Lambda $ equipped with a section
over the vertices.  Over each edge~$e$ (a component of~$\Lambda \setminus
\Lambda ^0$) the double cover is trivializable, and so by parallel transport
the trivializations at the two endpoints either agree~($0$) or
disagree~($1$).  This leads to an isomorphism
  \begin{equation}\label{eq:g47}
     \sD(\Lambda ,\Lambda ^0)\xrightarrow{\;\;\cong
     \;\;}\Map\bigl(\Edge(\Lambda ),\zt \bigr). 
  \end{equation}
The Hilbert space of the toric code is the complex vector space
  \begin{equation}\label{eq:g48}
     \sH = \Fun\bigl(\sD(\Lambda ,\Lambda ^0);\CC \bigr) 
  \end{equation}
of complex-valued functions on the finite set~\eqref{eq:g46}.  It carries the
hermitian inner product that makes the basis of $\delta $-functions
orthonormal.  Using~\eqref{eq:g47} we see the tensor product decomposition 
  \begin{equation}\label{eq:11}
     \sH \cong \bigotimes\limits_{e\in \Edge(\Lambda )} \sH_e, 
  \end{equation}
where $\sH_e = \Fun\bigl(\sD(e,\partial e);\CC \bigr)$ and the tensor product
is over~$\CC$.
 
A vertex $v\in \Lambda ^0$ determines the automorphism
  \begin{equation}\label{eq:g49}
     \varphi _v\:\sD(\Lambda ,\Lambda ^0)\longrightarrow \sD(\Lambda ,\Lambda
     ^0) 
  \end{equation}
that flips the section at the vertex~$v$.  A face~$f$ (a component
of~$Y\setminus \Lambda $) has a boundary that is homeomorphic to a circle, and
a double cover of a circle is either trivial~(0) or nontrivial~(1).  Let
  \begin{equation}\label{eq:g50}
     h_f\:\sD(\Lambda ,\Lambda ^0)\longrightarrow \{0,1\} 
  \end{equation}
be the function that returns the isomorphism type of the restriction of the
double cover to~$\partial f$.  Set
  \begin{equation}\label{eq:g51}
  \begin{aligned} 
      H_v\psi &= \frac 12(\psi -\varphi _v^*\psi ),\qquad \psi
     \in \sH \\ H_f\psi &= h_f\psi 
  \end{aligned}
  \end{equation}
Each operator~$H_v,H_f$ is a self-adjoint projection.  The kernel of~$H_v$
consists of functions invariant under flip of the section at~$v$, and the
kernel of~$H_f$ consists of functions supported at double covers that extend
over the face~$f$.  It is a nice exercise to verify that any two operators
in~\eqref{eq:g51} commute.
 
The Hamiltonian of the toric code is the sum these of commuting projection
operators:
  \begin{equation}\label{eq:g52}
     H=\sum\limits_{v}H_v \;+\;\sum\limits_{f} H_f. 
  \end{equation}
It follows that the spectrum of~$H$ is contained in the set~$\ZZ^{\ge0}\subset
\RR$ of nonnegative integers.  Furthermore, the kernel of~$H$ consists of
functions supported on double covers that extend over~$Y$.  More precisely, in
the diagram
  \begin{equation}\label{eq:g53}
     \sD(\Lambda ,\Lambda ^0)\xrightarrow{\;\;\pi \;\;} \sD(\Lambda
     )\xleftarrow{\;\;r\;\;} \sD(Y) 
  \end{equation}
the map~$\pi $ forgets the section~$s$, and $r$~ restricts a double cover
over~$Y$ to~$\Lambda \subset Y$.  The map~ $r$ is injective.  Define an
embedding
  \begin{equation}\label{eq:g54}
     \Map\bigl(\sD(Y),\CC \bigr)\longhookrightarrow \sH 
  \end{equation}
that takes a function on~$\sD(Y)$, extends it by zero to a function
on~$\sD(\Lambda )$, and then pulls back via~$\pi $ to a function
on~$\sD(\Lambda ,\Lambda ^0)$.  The descriptions of $\ker H_v$ and $\ker H_f$
show that the image of~\eqref{eq:g54} is the vacuum space~$\ker H$.

  \begin{remark}[]\label{thm:g21}
 \ 
 \begin{enumerate}[label=\textnormal{(\arabic*)}]

 \item The toric code is gapped in a strong sense: for any $\Lambda \subset Y$
the Hamiltonian has a spectral gap that is independent of~$\Lambda $: the gap
is bounded below by~1 since the spectrum for any~$\Lambda $ is contained in
$\ZZ^{\ge0}\subset \RR$. 
 
 \item One can formulate the toric code on any CW complex.  In other words, in
this model one can take the discrete structure on space~$Y$ to be a
CW~structure.  At the end of~\S\ref{sec:2} an alternative emerges: the discrete
structure can be two disjoint finite subsets of~$Y$.

 \end{enumerate}
  \end{remark}

   \section{Lattice systems from 3-dimensional topological field theories}\label{sec:2}

We introduce a new class of lattice models.  These models are gapped and have
commuting projector Hamiltonians.  They enjoy the two locality properties
highlighted for quantum field theory in the introduction.  The starting point
is a topological field theory, which is then the low energy limit of the gapped
lattice model.  Our exposition is for the special case of $2+1$-dimensional
models and codimension~2 defects; there are clear generalizations to explore in
other spacetime dimensions and with other dimensions of defects.  At the end of
this section we explain how to recover the toric code, though we emphasize that
the interest here is more in the newer models, such as the realization of
arbitrary 3-dimensional Chern-Simons theories via lattice models.
 
Let $\bord{1,2,3}(\sF)$ denote the bordism 2-category of 1-, 2-, and
3-dimensional manifolds equipped with a section of a sheaf~$\sF$ of
background fields, let $\Cat_{\CC}$ denote a 2-category of categories, and let
  \begin{equation}\label{eq:1}
     F\:\bord{1,2,3}(\sF)\longrightarrow \Cat_{\CC} 
  \end{equation}
be a \emph{Reshetikhin--Turaev} topological field
theory~\cite{RT1,RT2}.\footnote{We defer to the notes~\cite{F} and the
references therein for details of and motivation for the setup of topological
field theory, and to~\cite{FT2} for an axiomatization of Reshetikhin--Turaev
theories.}  A prime example is {\it Chern--Simons theory\/}, specified by a
compact Lie group and a level.  In general, a Reshetikhin--Turaev theory is
constructed from a modular tensor category
  \begin{equation}\label{eq:2}
     B = F(\cir)
  \end{equation}
and a lift of its central charge from~$\QQ/8\ZZ$ to~$\QQ/24\ZZ$.  (`$B$'~is for
`braided'.)  The sheaf of background fields is
  \begin{equation}\label{eq:3}
     \sF = \bigl\{\textnormal{orientation, $p_1$-structure}\bigr\} =
     \bigl\{(w_1,p_1)-\textnormal{structure}\bigr\}. 
  \end{equation}
In other words, the theory is formulated on oriented manifolds equipped with a
$p_1$-structure.  See~\cite[Appendix~B]{BHMV} for a definition of
$p_1$-structures; Atiyah's 2-framings~\cite{A} are closely related.
\emph{Turaev--Viro} theories~\cite{TV} are a special class of
Reshetikhin--Turaev theories, and they do not require a $p_1$-structure.
Finite gauge theories are a prime example of Turaev--Viro theories.
Turaev--Viro theories admit topological boundary theories, whereas the general
Reshetikhin--Turaev theory does not~\cite{FT2}.

Topological field theory is enriched by placing \emph{defects} on objects and
morphisms in the bordism category and then evaluating appropriately.  A defect
has support on a stratified manifold.  Boundary theories and domain walls are
special defects.  Defects in topological field theory have a long history;
see~\cite[\S\S2.\{4,5\}]{FMT} for a recent account.

To construct our lattice model, fix a finite set~$D$ of objects in~$B$, each of
the form~$1+b$ for some~$b\in B$.  Here `$1$'~is the tensor unit object in~$B$.
In the simplest case, $D$~has a single element.

  \begin{figure}[ht]
  \centering
  \includegraphics[scale=1.6]{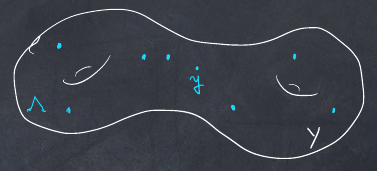}
  \vskip -.5pc
  \caption{A closed $(w_1,p_1)$-surface with codimension~2 defects (normal
  framing not drawn)}\label{fig:2}
  \end{figure}
  
Let $Y$~be a closed 2-manifold equipped with a $(w_1,p_1)$-structure, and fix a
finite subset~$\Delta \subset Y$ as well as a framing\footnote{This is the
usual normal framing of a Wilson loop in Chern-Simons theory~\cite{W}.  Since
$\Delta $~is a finite set of points, this is a framing of the tangent space
to~$Y$ at each point~$y\in \Delta $.} of the normal bundle to~$\Delta \subset
Y$; see~\autoref{fig:2}.  A defect at~$y\in Y$ is an element of the vector
space $\Hom\bigl(1,F(\cir(y)) \bigr)$, where $\Hom$~is taken in the
2-category~$\Cat_{\CC}$; $1$~is the tensor unit, which is the
category~$\Vect_{\CC}$ of complex vector spaces; and $S^1(y)$~is a linking
circle to~$y$.  Use the framing to identify~ $\cir(y)$ with~$\cir$ and so
identify $F(\cir(y))$~with the modular tensor category~$B$.  Fix a map $\delta
\:\Delta \to D$ and use it to place a $D$-defect at each point of~$\Delta $.
The value of the topological field theory~$F$ on~$Y$ with its
$(w_1,p_1)$-structure and collection of defects is the vector
space~$\sH(Y,\Delta ,\delta )$ of states of a discrete quantum system.

  \begin{figure}[ht]
  \centering
  \includegraphics[scale=1.0]{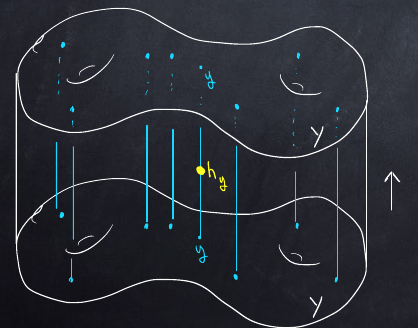}
  \vskip -.5pc
  \caption{The 3-dimensional bordism $(Y,\Delta ,\delta )\to (Y,\Delta ,\delta
  )$ that computes~$H_y$}\label{fig:3}
  \end{figure}

Define a commuting projector Hamiltonian on ~$\sH(Y,\Delta ,\delta )$ as
follows.  For each $y\in \Delta $ consider the Cartesian product $[0,1]\times
Y$ with codimension two defects~$\delta (y')$ supported at $[0,1]\times
\{y'\}$, $y'\in \Delta $, and an additional embedded point defect at
$\{1/2\}\times \{y\}$; see \autoref{fig:3}.  The latter is an element of
$\End_B(1+b)$, where $1+b\in B$ is the defect~$\delta (y)$ at~$y\in Y$, and we
take it to be the endomorphism given by the matrix
  \begin{equation}\label{eq:4}
     h_y = \begin{pmatrix} 0&0\\ 0&\id_b \end{pmatrix} .
  \end{equation}
Let 
  \begin{equation}\label{eq:5}
     H_y\: \sH(Y,\Delta ,\delta )\longrightarrow \sH(Y,\Delta ,\delta )
  \end{equation}
be the value of~$F$ on the resulting bordism.  Elementary standard arguments in
topological field theory prove that $H_y$~is a projection operator and that the
operators $\{H_y\}_{y\in \Delta }$ pairwise commute.  Furthermore, there is a
direct sum decomposition 
  \begin{equation}\label{eq:7}
     \sH(Y,\Delta ,\delta ) = \sH(Y,\Delta ,\delta ')\;\oplus \; \sH(Y,\Delta
     ,\delta ''),
  \end{equation}
where in the first summand we replace the defect at~$y$ with the transparent
defect (labeled~`$1$') and in the second summand we replace the defect~ $1+b$
at~$y$ with the defect~$b$.  This is the eigenspace decomposition of~$H_y$.
The first summand is the kernel and $H_y$~acts as the identity operator on the
second summand.  Define the Hamiltonian
  \begin{equation}\label{eq:6}
     H = \sum\limits_{y\in \Delta }H_y. 
  \end{equation}
Then $\spec H\subset \Znn$ and the kernel of~$H$ is $F(Y)\subset \sH(Y,\Delta
,\delta )$, the vector space of the underlying topological field theory.

  \begin{figure}[ht]
  \centering
  \includegraphics[scale=1.6]{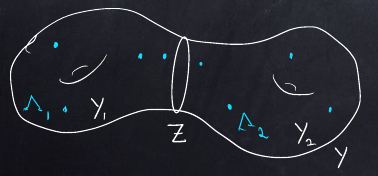}
  \vskip -.5pc
  \caption{Locality of the state space; see~\eqref{eq:9}}\label{fig:13}
  \end{figure}

  \begin{remark}[]\label{thm:1}
 \ 
 \begin{enumerate}[label=\textnormal{(\arabic*)}]

 \item If we split $Y=Y_1\cup _{Z}Y_2$ along a circle, and correspondingly
split $\Delta =\Delta _1\cup \Delta _2$ and $\delta =\delta _1\cup \delta _2$,
as depicted in \autoref{fig:13}, then the state space $\sH(Y,\Delta ,\delta )$
also decomposes.  Namely, the value $\sH(Y_i,\Delta _i,\delta _i)$ of~$F$ on
each surface with boundary is an object in~$B$, and $\sH(Y,\Delta ,\delta )$ is
the Hom space between those objects.  Let $\sO$ denote the set of isomorphism
classes of simple objects in~$B$.  The semisimple category~ $B$ can be
identified as the category of vector bundles over~$\sO$.  If $\sH_i\to \sO$ is
the vector bundle $F(Y_i,\Delta _i,\delta _i)$, then
  \begin{equation}\label{eq:9}
     \sH(Y,\Delta ,\delta ) \cong \bigoplus\limits_{a\in \sO}\;
     (\sH_1)_a\otimes \mstrut _{\CC} (\sH_2)_{a^\vee}, 
  \end{equation}
where $a^{\vee}$ is the dual object to~$a$.

 \item Different choices of defect set~$D$ give different gapped lattice
systems with the same low energy topological field theory.

  \begin{figure}[ht]
  \centering
  \includegraphics[scale=1.0]{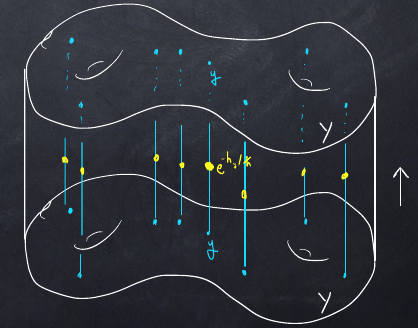}
  \vskip -.5pc
  \caption{The 3-dimensional bordism that computes $e^{-\tau H/\hbar}$}\label{fig:4}
  \end{figure}

 \item The Hamiltonian operator is a sum of evaluations of~$F$ on different
objects of the bordism category (allowing defects).  On the other hand, there
is a single evaluation of~$F$ that computes the evolution for imaginary
time~$\tau $, as depicted in \autoref{fig:4}: place embedded point defects at
$\{1/2\}\times \{y\}$ for \emph{all} $y\in \Delta $, and replace~\eqref{eq:4}
by the exponential $e^{-\tau h_y/\hbar}$.

 \end{enumerate}
  \end{remark}

  \begin{figure}[ht]
  \centering
  \includegraphics[scale=1.6]{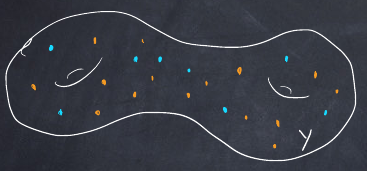}
  \vskip -.5pc
  \caption{A closed surface with two species $\color{cy}1+e$ and
  $\color{or}1+m$ of point defects}\label{fig:5}
  \end{figure}

\begin{figure}[ht]
  \centering
  \includegraphics[scale=1.6]{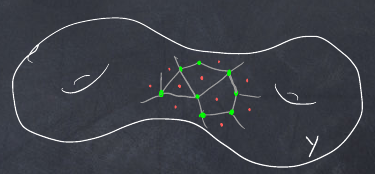}
  \vskip -.5pc
  \caption{Construction of $(Y,\Delta ,\delta )$ from $(Y,\Lambda
  )$}\label{fig:7}
  \end{figure}

We conclude by explaining how to recover the toric code as a special case of
our construction.  Let $G$~be the finite cyclic group of order two, and let
$F=\FG$~be 3-dimensional finite $G$-gauge theory.  The modular tensor
category~$B$ is the category $\Vect_G(G)$ of conjugation-equivariant vector
bundles over~$G$.  There are 4~simple objects $1,e,m,\psi $.  The regular
representation supported at~$e\in G$ is $1+e$ and the regular representation
supported at the non-identity element of~$G$ is~$m+\psi $.  The fusion rules
are $e^2 = m^2 =1$ and $\psi =em$.  Choose $D=\{1+e,\, 1+m\}$.  Our
construction gives a lattice model on surfaces with two species of defects, as
depicted in \autoref{fig:5}.  In other words, in this model the discrete
structure of space is a finite set of points ``colored'' with a defect in~$D$.
For a surface equipped with an embedded graph, as in \autoref{fig:1}, place a
defect $1+e$ at each vertex and a defect $1+m$ in each face, as in
\autoref{fig:7}.  (We explain the change of color shortly.)  We claim that this
model is the toric code.

\begin{figure}[ht]
  \centering
  \includegraphics[scale=1.6]{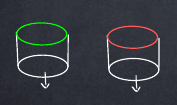}
  \vskip -.5pc
  \caption{A cylinder with Dirichlet/Neumann (green/red) boundary
  theory at one end}\label{fig:6}
  \end{figure}

To verify this claim, use the extension 
  \begin{equation}\label{eq:8}
     \tF\:\bord{0,1,2,3}\longrightarrow \Cat^{\otimes }_{\CC} 
  \end{equation}
of $G$-gauge theory to a fully local theory of unoriented manifolds with
codomain the 3-category of tensor categories.  This theory assigns the tensor
category $\Vect[G]$ to a point, and the modular tensor category~$B$ is its
Drinfeld center.  The theory~$\tF$ admits two topological boundary theories,
which we call Dirichlet (green) and Neumann (red).  There are semiclassical
descriptions.  The theory~$\tF$ sums over principal $G$-bundles.  Also sum over
trivializations of the $G$-bundle on boundary components colored with the
Dirichlet boundary theory; do nothing additional on boundary components colored
with the Neumann boundary theory.  We claim that the bordisms depicted in
\autoref{fig:6} evaluate to $1+e$ and $1+m$, respectively.  One way to see this
is to perform the finite path integral as in~\cite[Appendix]{FMT}.  This
amounts to a push-pull of vector bundles over the correspondences
  \begin{equation}\label{eq:10}
     \begin{gathered} \xymatrix{&\ast\ar[ld]\ar[rd] \\ \ast&& G\gpd
     G}\qquad\quad \qquad  
     \xymatrix{&G\gpd G\ar[ld]\ar[rd]^{\id} \\ \ast&& G\gpd G} \end{gathered} 
  \end{equation}

\begin{figure}[ht]
  \centering
  \includegraphics[scale=1.6]{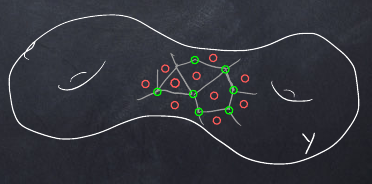}
  \vskip -.5pc
  \caption{Defects from boundary theories and simple locality in the toric code}\label{fig:8}
  \end{figure}

Quite generally, boundary theories produce defects supported on arbitrary
stratified manifolds.  Apply this to points to redraw \autoref{fig:7} as a
surface with point defects from boundary theories.  In \autoref{fig:8} we have
opted instead to draw the equivalent picture with small disks excised around
each defect, and we have colored the resulting boundary circles with Dirichlet
and Neumann boundary theories.  The semiclassical description of $G$-gauge
theory with Dirichlet and Neumann boundary theories reproduces precisely the
description of the toric code given in~\S\ref{sec:1}.

  \begin{remark}[]\label{thm:2}
 Although the general form~\eqref{eq:9} of locality in our models is a direct
sum of tensor products over~$\CC$, the toric code exhibits locality as a simple
tensor product~\eqref{eq:11} over~$\CC$.  The semiclassical argument we gave
above is specific to the toric code.  Here is a sketch of a quantum argument
that applies to all Turaev--Viro theories that admit a Neumann boundary theory.
Such theories are defined by finite dimensional semisimple Hopf
algebras~\cite[\S8B2]{FT1}.

  \begin{figure}[ht]
  \centering
  \includegraphics[scale=1.2]{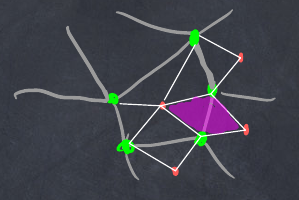}
  \vskip -.5pc
  \caption{Simpler locality along edges in Hopf algebra models}\label{fig:14}
  \end{figure}

\autoref{fig:14} depicts a small piece of a surface~$Y$ with an embedded
graph~$\Lambda $, as in \autoref{fig:7}.  A Dirichlet point defect is placed at
each vertex and a Neumann point defect is placed in the interior of each face.
For each edge consider the shaded region in \autoref{fig:14}.  The sides of
this quadrilateral are connected curves with one Dirichlet boundary and one
Neumann boundary.  Each evaluates to the category~$\Vect$, hence the entire
figure evaluates to a single vector space.  Upon gluing over sides of these
quadrilaterals we obtain a simple tensor product decomposition of~$\sH(Y,\Delta
,\delta )$ as a tensor product over~$\CC$ indexed by the edges of the
graph~$\Lambda $.  However, the argument as stated is too naive: to execute the
gluing in the bordism category one must blow up the figure along all sides and
vertices, as in~\cite{D}.
  \end{remark}

   \section{The Levin--Wen model}\label{sec:4}

The Levin--Wen lattice model, which generalizes the toric code, was introduced
in~\cite{LW}.  Our treatment here is inspired by~\cite{KK}, who go further and
construct lattice realizations of domain walls and topological boundary
theories.
 
Fix a fusion category~$C$, which we assume to be spherical and unitary.  It
determines a fully local topological field theory
  \begin{equation}\label{eq:15}
     \TC\:\bord{0,1,2,3}(\textnormal{orientation})\longrightarrow
     \Cat^{\otimes }_{\CC}
  \end{equation}
of oriented manifolds that satisfies $\TC(\pt)=C$: this is the associated
\emph{Turaev--Viro} theory~\cite{TV}.  Let $\rho $~be the Dirichlet boundary
theory, which is built from the regular $C$-module;
see~\cite[Definition~3.2]{FMT}.  The Levin--Wen model is a gapped lattice model
whose low energy theory is~$\TC$.  The construction uses point defects and the
Dirichlet boundary theory, as in~\S\ref{sec:2}, but in addition to these
defects one cuts out large open subsets of a surface as well.  The construction
bears some similarity to~\cite[\S7]{FT1}.  Applied to the fusion
category~$C=\Vect[G]$ we obtain yet another realization of the toric code.

  \begin{figure}[ht]
  \centering
  \includegraphics[scale=1.6]{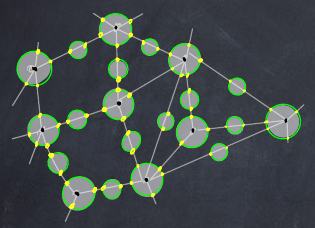}
  \vskip -.5pc
  \caption{A disjoint union of disks over vertices and edges}\label{fig:17}
  \end{figure}

Let $(Y,\Lambda )$ be a closed oriented surface~$Y$ with an embedded
graph~$\Lambda $.  We use notions of vertices, edges, and faces as
in~\S\ref{sec:1}.  Orient each edge of~$\Lambda $.  The state space is the
value of~$\TC$ on a disjoint union of disks about the vertices and interior
points of edges, as depicted in \autoref{fig:17}.  The boundaries of each disk
are colored with the Dirichlet boundary theory~$\rho $ (green), and there is
are embedded point defects (yellow) at the intersection points of disk
boundaries and edges.  The boundary circles inherit an orientation from the
orientation of~$Y$, or more precisely its restriction to an orientation of each
disk.  The link of an embedded point in the boundary circle is then an oriented
interval with Dirichlet ends, and $\TC$~ evaluates it to be the category
underlying~$C$.  Therefore, the embedded point defect is an object of~$C$.  Let
$\{c_i\}$ be a set of representative simple objects of~$C$, and define
  \begin{equation}\label{eq:16}
     c = c_1 + c_2 + \dots + c_N
  \end{equation}
To each (yellow) point of a disk with edge oriented out of the disk assign the
defect~$c$; if the edge is oriented into the disk assign the defect~$c\dual$.
Observe that $c\dual\cong c$, but there is no canonical isomorphism.  The disk
on an edge, with two point defects, evaluates to
  \begin{equation}\label{eq:17}
     \Hom\mstrut_C(1,c\otimes c\dual)=\bigoplus\limits_{i,j} \Hom\mstrut_C(1,c\mstrut
     _i\otimes c_j\dual) = \bigoplus\limits_{i} \Hom\mstrut_C(1,c\mstrut
     _i\otimes c_i\dual)= \CC^N.
  \end{equation}
(These isomorphisms are canonical, hence the $=$~signs.)  The disk at a
vertex~$v$, if it has $m_v^+$~outgoing edges and $m_v^-$~incoming edges,
evaluates to
  \begin{multline}\label{eq:18}
    \phantom{MMM}\Hom\mstrut_C(1,\underbrace{c\otimes \cdots\otimes
     c}_{\textnormal{$m_v^+$ times}}\otimes \underbrace{c\dual\otimes
     \cdots\otimes c\dual}_{\textnormal{$m_v^-$ times}}) \\[9pt] =
     \bigoplus\limits_{\substack{\phantom{M}i_1,\dots ,i_{m_v^+} \\
     \phantom{M}j_1,\dots ,j_{m_v^-}}} \Hom\mstrut_C(1,c\mstrut _{i_i}\otimes \cdots
     \otimes 
     c\mstrut _{i_{m_v^+}}\otimes c_{j_i}\dual\otimes \cdots \otimes
     c_{j_{m_v^-}}\dual).\phantom{MM}
  \end{multline}
The disjoint union of disks evaluates to a tensor product over~$\CC$ of the
vector spaces~\eqref{eq:17} and~\eqref{eq:18}: 
  \begin{equation}\label{eq:19}
     \bigotimes\limits_{v} \Hom_{C}(1,c^{\otimes m_v^+}\otimes
     (c\dual)^{\otimes m_v^-})\;\otimes\; \bigotimes\limits_{e}
     \Hom\mstrut_C(1,c\otimes c\dual).
  \end{equation}
This matches the state space of the Levin--Wen model (up to isomorphism).

  \begin{figure}[ht]
  \centering
  \includegraphics[scale=.7]{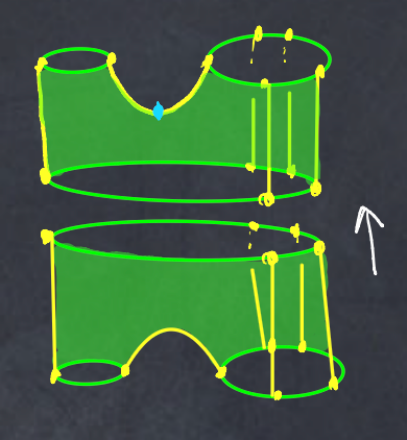}
  \vskip -.5pc
  \caption{The operator $P_{v,e}$}\label{fig:18}
  \end{figure}

\begin{figure}[ht]
  \centering
  \includegraphics[scale=.7]{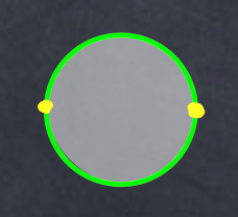}
  \vskip -.5pc
  \caption{The link of the embedded point defect in \autoref{fig:18}}\label{fig:25}
  \end{figure}

As in the toric code, the Hamiltonian of the Levin--Wen model is a sum of terms
attached to vertices and faces.  For each vertex~$v$ and abutting edge~$e$
consider the bordism depicted in \autoref{fig:18}.  This is a 3-manifold with
corners, it has two components, each component has three boundary disks, the
``sides'' are colored with~$\rho $, and there are defect curves on the
boundary, each ``labeled'' by~$c$.  In addition, there is an embedded point
defect in one of these $c$-curves, drawn in cyan.  The link of that point is
the disk depicted in \autoref{fig:25}, which evaluates to the vector
space 
  \begin{equation}\label{eq:23}
     \End\mstrut_C(c) = \bigoplus\limits_{i} \End\mstrut_C(c_i) = \bigoplus\limits_{i} \CC;
  \end{equation}
see \cite[Figure~7]{FT1}.  We choose the defect~$p$ to have value $1/\dim(c_i)$
on the $i^{\textnormal{th}}$~component:   
  \begin{equation}\label{eq:25}
     p = \bigoplus\limits_{i} \frac{1}{\dim(c_i)}. 
  \end{equation}
Note that the dimension of the simple object~$c_i$ is nonzero
by~\cite[Corollary~5.15]{ENO}.  The value of~$\TC$ applied to the bordism of
\autoref{fig:18} is a linear endomorphism of the tensor product of the vector
spaces associated to the two lower boundary disks, i.e., of the tensor product
of the vector space~\eqref{eq:17} attached to~$e$ and the vector space
~\eqref{eq:18} attached to~$v$.  Extend to a linear endomorphism~$P_{v,e}$
of~\eqref{eq:19} by acting as the identity on all other tensor factors.

  \begin{figure}[ht]
  \centering
  \includegraphics[scale=.7]{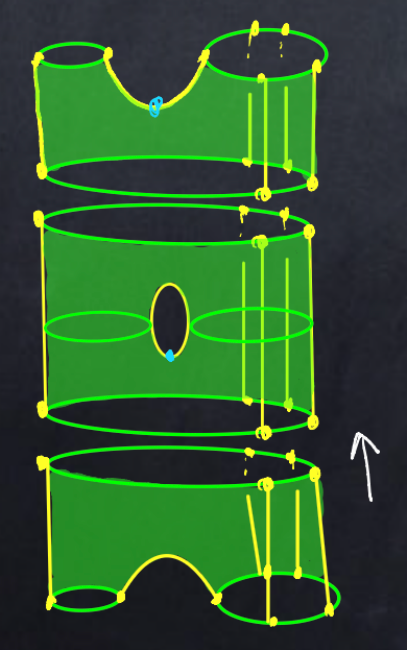}
  \vskip -.5pc
  \caption{The square $P_{v,e}\circ P_{v,e}$}\label{fig:19}
  \end{figure}

\begin{figure}[ht]
  \centering
  \includegraphics[scale=.6]{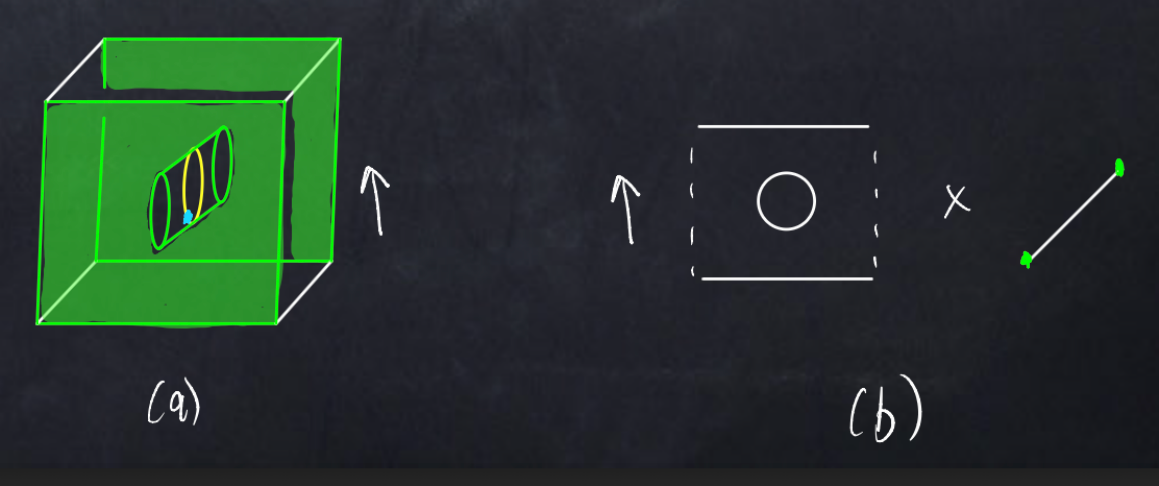}
  \vskip -.5pc
  \caption{\parbox[t]{30pc}{
  (a) The central component of \autoref{fig:19} chopped on the sides
  \\(b)~a neighborhood of the central hole}}\label{fig:20}
  \end{figure}

We claim\footnote{The reader may want to verify this first in gauge theory
($C=\Vect[G]$) using the semiclassical construction as a sum over $G$-bundles,
though in this case $\dim(c)=1$ for all simple objects~$c$ so the more subtler
aspects of the computation are not in evidence.  The argument in the text is
quantum and applies generally.} that $P_{v,e}$~is a projection operator, i.e.,
$P_{v,e}\circ P_{v,e}=P_{v,e}$.  The squared operator is shown in
\autoref{fig:19}.  It suffices to show that the central component evaluates to
the identity operator.  That central component is isolated in
\autoref{fig:20}(a), where we have also chopped off the sides, including the
embedded point defects that run from bottom to top.  It suffices to show that
this bordism evaluates to the identity operator.  To be clear: this is a solid
rectangular box; the front and back are colored~$\rho $; there is an open solid
cylinder that has been bored out from front to back, with the resulting
cylindrical boundary colored~$\rho $; there is an embedded circle defect~$c$ in
the middle of that cylinder; and there is the point defect~\eqref{eq:25}
embedded in the circle defect.  If we cut out a neighborhood of the cylindrical
boundary component, then we are left with the Cartesian product drawn in
\autoref{fig:20}(b).\footnote{The vertical sides of the square are dotted in
the sense of~\cite[\S2.1.2]{FT2}.}  Consider the dimensional reduction of~$\TC$
along the interval with $\rho $-endpoints.  Since $\TC$~of that interval is the
underlying category~$C$, the dimensional reduction is the 2-dimensional theory
whose value on a point is~$C$.  This reduction is an oriented theory; the
Frobenius structure on the underlying category~$C$ was computed in
\cite[Proposition~6.18]{FT1}: for each object~$x\in C$ the trace~$\tau _x$ of
the identity endomorphism of~$x$ is
  \begin{equation}\label{eq:24}
     \tau _x(\id_x) = \dim(x). 
  \end{equation}
With proper ``arrows of time'' the top and bottom of the square in
\autoref{fig:20}(b) evaluate to the identity functor~$\id\mstrut_C$ on~$C$, hence the
entire bordism evaluates to an endomorphism of~$\id\mstrut_C$, i.e., to an element of
the Hochschild cohomology~ $HH^0(C)$.  It remains to show that it equals the
identity.

\begin{figure}[ht]
  \centering
  \includegraphics[scale=1.6]{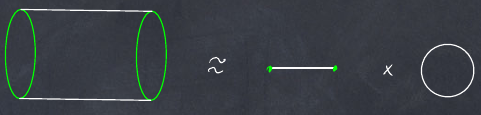}
  \vskip -.5pc
  \caption{The Hochschild (co)homology~$HH_0(C)\cong HH^0(C)$}\label{fig:26}
  \end{figure}

As a preliminary, consider the surface \autoref{fig:26}.  As we have noted
above, the dimensional reduction of~$\TC$ on the interval with $\rho
$-endpoints is the 2-dimensional theory whose value on a point is~$C$.  The
value of that theory on the circle is the Hochschild homology~$HH_0(C)$.  The
Frobenius structure~\eqref{eq:24} on~$C$ induces an isomorphism~$HH_0(C)\cong
HH^0(C)$ with the Hochschild cohomology.  

\begin{figure}[ht]
  \centering
  \includegraphics[scale=1.6]{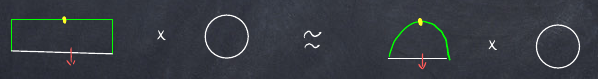}
  \vskip -.5pc
  \caption{The element of~$HH^0(C)$ produced in \autoref{fig:20}(a)}\label{fig:22}
  \end{figure}

Return now to the computation of the bordism in \autoref{fig:20}(a).  We
previously cut out a neighborhood of the central cylinder; that
neighborhood---minus the embedded cyan point defect---is depicted in
\autoref{fig:22} as a bordism whose evaluation is an element of ~$HH^0(C)$.  In
the figure the embedded yellow point defect is~\eqref{eq:16}.  Then,
tautologically, the first factor of the Cartesian product in \autoref{fig:22}
evaluates to $c\in C$.  The Cartesian product with the circle is the
application of the Hochschild homology functor~$HH_0(-)$, which maps $c\in C$
to $\id_c\in HH_0(C)$.  Under the canonical isomorphism
  \begin{equation}\label{eq:26}
     HH_0(C) = \bigoplus\limits_{i} \End\mstrut_C(c_i) = \bigoplus\limits_{i}
     \CC,
  \end{equation}
this is the vector $\oplus _i\,1$.  Apply the isomorphism with Hochschild
cohomology to obtain 
  \begin{equation}\label{eq:27}
     \oplus _i\,\dim(c_i)\;\in \; HH^0(C) = \bigoplus\limits_{i}
     \End\mstrut_C(c_i) = \bigoplus\limits_{i} \CC.
  \end{equation}
Finally, the embedded point defect~\eqref{eq:25} compensates for the Frobenius
trace.  In all, the central component of \autoref{fig:19} evaluates to the
identity, as claimed.

\begin{figure}[ht]
  \centering
  \includegraphics[scale=1.6]{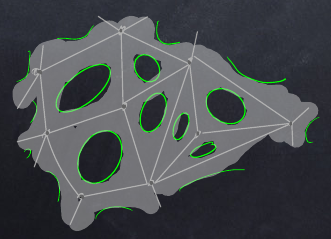}
  \vskip -.5pc
  \caption{The surface that evaluates to the kernel of $\sum\limits_{v}H_v$u}\label{fig:23}
  \end{figure}

For a fixed vertex~$v$ define a linear endomorphism~$H_v$ of~\eqref{eq:19} as 
  \begin{equation}\label{eq:21}
     H_v = \id - \prod\limits_{e}P_{v,e}, 
  \end{equation}
where the product is over all edges~$e$ that emanate from~$v$.  This is an
operator with spectrum~$\{0,1\}$.  From the construction, the kernel is the
evaluation of~$\TC$ on the union of disks obtained from \autoref{fig:17} by
joining the disk at the fixed vertex~$v$ to the disk at each edge~$e$ emanating
from~$v$; in the process eliminate the embedded yellow point defects on the
edge~$e$.  After repeating for each vertex~$v$ we obtain \autoref{fig:23}.

\begin{figure}[ht]
  \centering
  \includegraphics[scale=1.6]{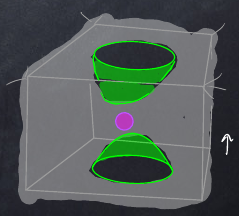}
  \vskip -.5pc
  \caption{The operator~$\id-H_f$, including the purple antisphere}\label{fig:24}
  \end{figure}

The Hamiltonian of the Levin--Wen model is the sum of~\eqref{eq:21} over the
vertices and the sum over faces~$f$ of a projection operator~$H_f$.  The
construction of~$H_f$ follows~\cite[\S7A]{FT1}.  Namely, the vector
space~$\TC(S^2)$ is one dimensional and has a canonical element~$\TC(D^3)$, the
value of the theory on the closed 3-disk, so we identify $\TC(S^2)=\CC$.  In
\cite[(6.23)]{FT1} we evaluate~$\TC$ on $[0,1]\times S^2$ with $\{0\}\times
S^2$ painted~$\rho $; the value is $d(C)$, the categorical dimension of~$C$,
which is the sum of the squares of the dimensions of the simple objects.  This
number is nonzero.  We invert it to define an ``antisphere'', which is pictured
in purple in \autoref{fig:24}.  The bordism in that picture defines an
operator~$\id-H_f$ on the vector space computed in \autoref{fig:23}; what is
not depicted is that on all other faces we use a Cartesian product, i.e., a
green cylinder running from bottom to top.  Extend the operator~$H_f$ to be the
identity on the complement of (the $\TC$-values of) \autoref{fig:23} in
\autoref{fig:17}.

The construction makes clear that the Levin--Wen Hamiltonian
  \begin{equation}\label{eq:22}
     H = \sum\limits_{v}H_v\;+\; \sum\limits_{f}H_f 
  \end{equation}
has spectrum contained in~$\Znn$ and the kernel is the vector space~$\TC(Y)$.

   \section{Reprise: a stat mech model in $1+1$ dimensions}\label{sec:3}

The models in~\S\ref{sec:2} have continuous time.  One can also use topological
field theory with defects to construct models with discrete time, that is, stat
mech models.  This was done for the $(1+1)$-dimensional Ising model
in~\cite{FT1}, so here we limit ourselves to a brief recapitulation and defer
to~\cite{FT1} for details and proofs.

  \begin{figure}[ht]
  \centering
  \includegraphics[scale=1.2]{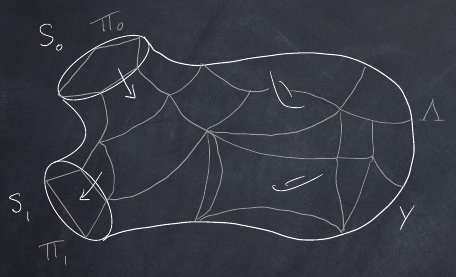}
  \vskip -.5pc
  \caption{A 2-dimensional bordism with embedded graphs}\label{fig:9}
  \end{figure}
 
The Ising model~\cite{C,ID} can be regarded as a 2-dimensional field theory on
manifolds equipped with an appropriate embedded graph.  A bordism $Y\:S_0\to
S_1$ of this type is depicted in \autoref{fig:9}, where $\Lambda \subset Y$ is
an embedded graph that restricts on each boundary~$S_i$ to an embedded
graph~$\Pi _i$.  Fix a finite group~$G$.  The model is defined by a finite path
integral over \emph{spins} $\sigma \:\Lambda ^0\to G$.  As in~\eqref{eq:10} we
compute using a correspondence of function spaces
  \begin{equation}\label{eq:12}
     \begin{gathered} \xymatrix{&G^{\Lambda ^0}\ar[ld]_{r_0}\ar[rd]^{r_1} \\
     G^{(\Pi _0)^0}&& G^{(\Pi _0)^0}} \end{gathered} 
  \end{equation}
The maps $r_0,r_1$ restrict a spin on~$Y$ to the incoming and outgoing boundary
components, respectively.  Now rather than a simple pull-push, the map on
functions includes a kernel $K\:G^{\Lambda ^0}\to \Rp$: it is $(r_1)_*\circ
K\circ (r_0)^*$.  The kernel depends on a weight function $\theta \:G\to \Rp$
which is required to satisfy, among other conditions, that $\theta (g\inv
)=\theta (g)$.  Then for each edge~$e$ in~$\Lambda $ the value of~$\theta $ on
the ratio of spins at the endpoints, denoted $\theta \bigl(\sigma (\delta
e)\bigr)$, is well-defined.  The kernel is
  \begin{equation}\label{eq:13}
     K(\sigma ) = \prod\limits_{e}\theta \bigl(\sigma (\delta e)\bigr). 
  \end{equation}
On a closed surface we obtain the usual Ising partition function. 
 
  \begin{figure}[ht]
  \centering
  \includegraphics[scale=1.2]{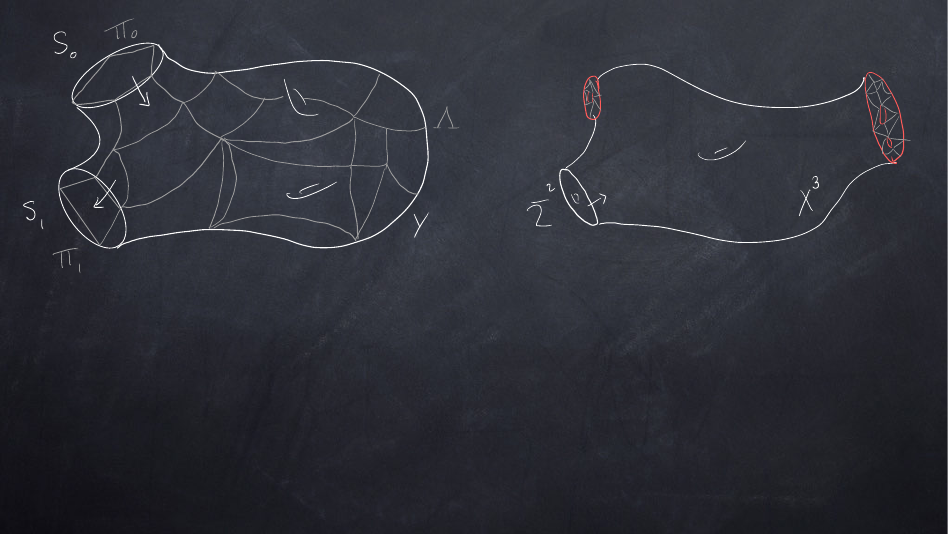}
  \vskip -.5pc
  \caption{Ising as a boundary of 3-dimensional finite gauge theory}\label{fig:10}
  \end{figure}

\begin{figure}[ht]
  \centering
  \includegraphics[scale=1.6]{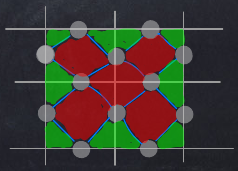}
  \vskip -.5pc
  \caption{A piece of surface colored with $\rho ,\epsilon ,\delta $}\label{fig:15}
  \end{figure}

\begin{figure}[ht]
  \centering
  \includegraphics[scale=1.2]{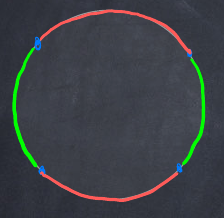}
  \vskip -.5pc
  \caption{The incoming disk~$D_e$}\label{fig:11}
  \end{figure}

The group~$G$ acts as a symmetry of the Ising model: left translate all spins
simultaneously.  Thus Ising can be realized as a boundary theory of the
3-dimensional finite gauge theory~$\tF$ that appeared in~\eqref{eq:8}; see
\autoref{fig:10}.  (More precisely, the finite gauge theory quiche acts, as
in~\cite{FMT}.)  We use three special defects in~$\tF$: the Dirichlet boundary
theory~$\rho $, the Neumann boundary theory~$\epsilon $, and the unique
irreducible interface~$\delta $ between them.  \autoref{fig:15} illustrates how
we use these defects.  Depicted is a piece of surface~$Y$ with an embedded
graph~$\Lambda $ (in gray).  Small disks are excised around the middle of each
edge.  The remainder is ``colored'' with the Dirichlet theory~$\rho $ (green),
the Neumann theory~$\epsilon $ (red), and the interface~$\delta $ (blue)
between them.  This appears on the boundary $\{0\}\times Y$ of the bordism
$X=[0,1]\times Y$.  For convenience, color the boundary $\{1\}\times Y$ with the
Dirichlet theory as well.  Then we read this 3-manifold with partially colored
boundary as a bordism 
  \begin{equation}\label{eq:14}
     X\:\bigtimes_{e}D_e\to \emptyset ^2, 
  \end{equation}
where the Cartesian product is over the edges of~$\Lambda $ and the incoming
disk~$D_e$ is pictured in \autoref{fig:11}; $\emptyset ^2$~is the empty set as
a 2-manifold.  Gauge theory~$\tF$ evaluates on the disk~$D_e$ with colored
boundary to a vector space.  It is straightforward to compute this vector space
in the semiclassical path integral.  Namely, the groupoid of $G$-bundles
on~$D_e$, trivialized over the Dirichlet boundary, is equivalent to~$G$.
Therefore, the quantization---the value of~$\tF$---is the vector space of
functions.  It is now an easy matter to see that if we plug in the weight
function~$\theta $ at each~$\tF(D_e)$, then the theory~$\tF$ evaluates on~$X$
to the Ising partition function.

This construction of the Ising model within topological field theory has many
consequences.  First, an application of the cobordism hypothesis \emph{proves}
Kramers--Wannier duality, in fact as a boundary consequence of 3-dimensional
finite electromagnetic duality.  Second, one finds duals to Ising models for
nonabelian~$G$.  These duals are constructed using the cobordism hypothesis.
Third, one can make predictions for the gapped phases of these models based on
a classification of \emph{topological} boundary theories of~$\tF$.  We can also
incorporate the Kramers--Wannier duality defect~\cite{FFRS,AMF,CCHLS,KOZ} into
this picture.  We refer to~\cite{FT1} for these consequences and more.

 \bigskip\bigskip
 \bibliographystyle{hyperamsalpha} 
\newcommand{\etalchar}[1]{$^{#1}$}
\providecommand{\bysame}{\leavevmode\hbox to3em{\hrulefill}\thinspace}
\providecommand{\MR}{\relax\ifhmode\unskip\space\fi MR }
\providecommand{\MRhref}[2]{%
  \href{http://www.ams.org/mathscinet-getitem?mr=#1}{#2}
}
\providecommand{\href}[2]{#2}

  \end{document}